\documentclass{iau}
\usepackage{graphicx}
\usepackage{enumitem}

\title[Galactic warp signal in Gaia DR1 proper motions] 
{Search for Galactic warp signal \\ in Gaia DR1 proper motions}

\author[E. Poggio et al.]   
{E. Poggio$^{1,2},$
 R. Drimmel$^2$,
 R. L. Smart$^{2,3}$,
 A. Spagna$^2$,
 M. G. Lattanzi$^2$}

\affiliation{$^1$Universit\`a di Torino, Dipartimento di Fisica, via P. Giuria 1, 10125 Torino, Italy \\ email: {\tt poggio@oato.inaf.it} \\[\affilskip]
$^2$Osservatorio Astrofisico di Torino, Istituto Nazionale di Astrofisica (INAF), Strada Osservatorio 20, 10025 Pino
Torinese, Italy \\
$^3$ School of Physics, Astronomy and Mathematics, University of Hertfordshire, College Lane, Hatfield AL10 9AB, UK\\
}

\pubyear{2017}
\volume{330}  
\jname{Astrometry and Astrophysics in the Gaia sky}
\editors{A. Recio-Blanco, P. de Laverny, A. G. A. Brown \& T. Prusti, eds.}
\begin{document}

\maketitle

\begin{abstract}
The nature and origin of the Galactic warp represent one of the open questions posed by Galactic evolution. Thanks to
Gaia high precision absolute astrometry, steps towards the understanding of the warp's dynamical nature can be made.
Indeed, proper motions for long-lived stable warp are expected to show measurable trends in the component vertical to
the galactic plane. Within this context, we search for the kinematic warp signal in the first \emph{Gaia} data release
(DR1). By analyzing distant spectroscopically-identified OB stars in the Hipparcos subset in \emph{Gaia} DR1, we find
that the kinematic trends cannot be explained by a simple model of a long-lived warp. We therefore discuss possible
scenarios for the interpretation of the obtained results. We also present current work in progress to select a larger
sample of OB star candidates from the \emph{Tycho-Gaia} Astrometric Solution (TGAS) subsample in DR1, and delineate the
points that we will be addressing in the near future.

\end{abstract}

\firstsection 
\section{Introduction}
The warp of the Milky Way is a well known feature of the outer disk, whose presence has been detected in the gas, dust and
stars. However, its dynamical nature - as well as the formation mechanism - continues to remain an unsolved mystery. In
the case of a long-lived static warp, systematic vertical motions would result in measurable trends in the stellar
proper motions (\cite[Smart \& Lattanzi 1996]{Smart:1996}), which become most significant toward the Galactic anti-center. We select OB stars since they can be
seen to large distances and are expected to trace the gaseous disk, in which the warp was originally detected. Given
the unprecedented astrometric precision, we first search for the warp kinematic signal the Hipparcos subset in the
first \emph{Gaia} data release (DR1, \cite[Gaia Collaboration 2016]{GaiaDR1}) (see \cite[Poggio et al. 2017,
hereafter Paper I]{PaperI}). Section \ref{HIPDR1} gives a brief overview of the obtained results and discusses
possible interpretations. Section \ref{TGASDR1} is dedicated to the selection of OB stars candidates in the
\emph{Tycho-Gaia} Astrometric Solution (TGAS) sample in DR1.

\section{Hipparcos subset in Gaia DR1 \label{HIPDR1}}

{\underline{\it Data selection}}. From the Hipparcos catalog (\cite[van Leeuwen 2008]{vanLeeuwen:2008}), we
spectroscopically select the OB3 stars with apparent magnitude $m_V < 8.5$ and parallax $\varpi <$ 2 mas, in order to
remove local structures.
The resulting selection contains 989 stars, among which 758 are present in the Hipparcos subsample in \emph{Gaia} DR1.
In the following, we present and discuss the results obtained with the smaller sample having \emph{Gaia} superior
astrometry.

{\underline{\it The model}}. The modelled spatial distribution consists of four major spiral arms
(\cite[Georgelin \& Georgelin 1976]{Georgelin:1976}, \cite[Taylor \& Cordes 1993]{Taylor:1993}) and one local arm, with a gaussian density profile in the Galactic plane
and an exponential vertical profile. Some spatial parameters are taken from the literature, while others are tuned to
reproduce the longitude, latitude and apparent magnitude distribution observed in the data (see Paper I for the
details). The kinematics is described by a simple model which includes solar motion from \cite{Schoenrich:2010},
Galactic rotation from \cite{Bland:2016} and velocity dispersions from \cite{Dehnen:1999}. Astrometric errors are
included in the model, together with the selection function of both Hipparcos catalogue and Hipparcos subset in
\emph{Gaia} DR1. Finally, the warp can be incorporated as a vertical displacement $z_W(R,\phi)$ in the $z$ spatial
coordinate, while its kinematic signal has a systematic offset $v_{z,W}(R,\phi)$ in the vertical velocities $v_z$.

\begin{figure}[tb]
\begin{center}
 \includegraphics[width=3.5in]{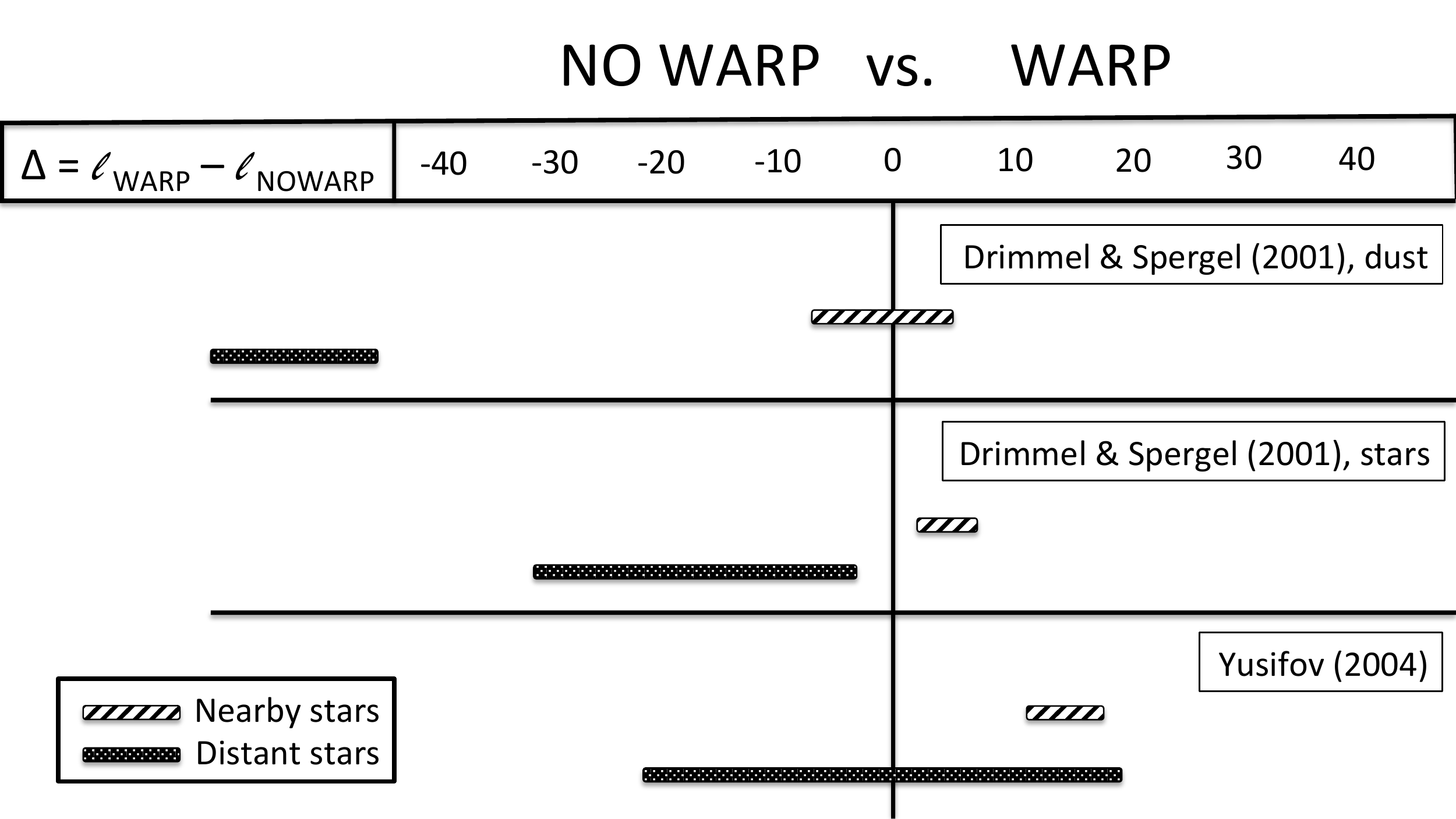}
 \caption{The difference of the loglikelihoods $\Delta = \ell_{WARP} - \ell_{NOWARP}$ (i.e. the likelihood ratio) for
 three different warp models according to the nearby ($1 < \varpi < 2$ mas) and distant ($\varpi < 1$ mas) stars of the sample.
 Positive (negative) values of $\Delta$ favour the warp (nowarp) model. The length of each bar is proportional to the
 uncertainty in $\Delta$, calculated using bootstrap resamples.}
   \label{results_hipdr1}
\end{center}
\end{figure}

{\underline{\it Results}}. Depending on the warp spatial parameters, different kinematic signals are expected. Here we
consider the three different sets of warp parameters, from \cite{Drimmel:2001} (both dust and stars) and
\cite{Yusifov:2004}. The kinematic signal predicted by each of them is compared to the alternative model $-$ the
\emph{nowarp} model, i.e. the absence of warp signal $-$, by constructing the expected probability distribution function
in the proper motions $\mu_b$ as a function of galactic longitude $l$. Figure \ref{results_hipdr1} summarizes our
results, showing the likelihood of the various warp models with respect to the no warp model for our TGAS-Hipparcos
dataset, divided into nearby ($1 < \varpi < 2$ mas) and distant ($\varpi < 1$ mas) stars. Our model for a long-lived
warp predicts that the kinematic signal becomes stronger for larger distances, while the data do not show any evidence
of warp signal for the most distant stars, consistent with the previous works of \cite{Smart:1998} and
\cite{Drimmel:2000}.

{\underline{\it Interpretations}}.
The absence of the warp kinematic signal in our dataset can be explained by several different scenarios. The first
possible explanation is that our sample of OB stars in the Hipparcos subset of \emph{Gaia} DR1 (approximately 0.5-3 kpc
from the Sun) is not sufficiently sampling the Galactic warp. Indeed, there is no consensus about where the warp
starts, although most studies find that the warp starts inside or close to the Solar circle (\cite[Momany et al.
2006]{Momany:2006}; \cite[Reyl\'{e} et al. 2009]{Reyle:2009}). Another possible interpretation is that the warp signal
is overwhelmed by other perturbations, such as vertical waves in the disk (\cite[G\'{o}mez et al. 2013]{Gomez:2013}).
Finally, it might be that our model of a long-lived stable warp is not appropriate, and additional effects should be
taken into account, like precession or an amplitude varying with time. To shed further light on the nature of the
Galactic warp it will be necessary to consider a larger dataset that samples a larger volumn of the Galactic disk.

\section{TGAS in Gaia DR1 \label{TGASDR1}}
\begin{figure}[tb]
\begin{center}
 \includegraphics[width=3.6in]{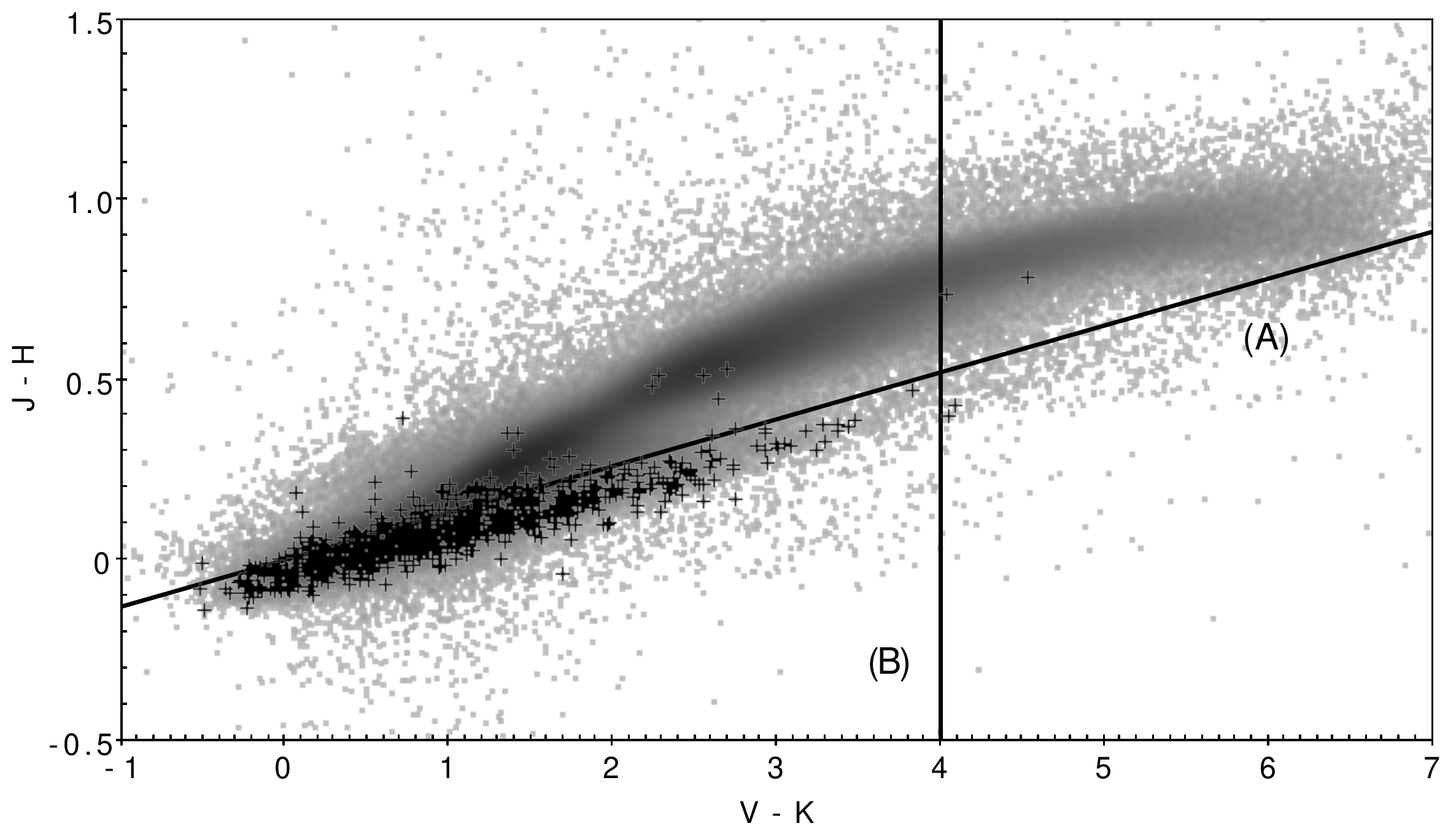}
 \caption{The TGAS stars with 2MASS and APASS photometry are represented by the grey dots, with the grey scale showing
their density. Known O-B3 stars from the Tycho-2 Spectral Type Catalogue (black crosses) are located in the bottom-left
part of the plot. Lines (A) and (B) are used for the selection process (see text).}
   \label{colcolsel}
\end{center}
\end{figure}
%

The selection of OB stars from the TGAS catalog is not trivial. Indeed, spectral classifications or parameter estimates
such as $\log(g)$ and $T_{\rm eff}$ are available in literature only for a small fraction of TGAS stars, and not for the
all sky. We therefore developed a selection criterium which combines astrometric with photometric measurements from the
2MASS (\cite[Skrutskie et al. 2006]{2MASS:2006}) and APASS (\cite[Henden et al. 2016]{APASS:2016}) surveys, available
for 1\,824\,237 TGAS stars. The first step of the method takes advantage of the fact that different stellar populations lie
in different regions of the color-color plot shown in Figure \ref{colcolsel}. Known O-B3 stars from the Tycho-2
Spectral Type Catalogue (\cite[Wright et al. 2003, black crosses]{Wright:2003}) are overplotted with the TGAS stars
(grey density map). Extinction moves the OB stars to the right, but separated from the majority of giants and stars on the
main-sequence. We begin by selecting all the TGAS stars below line (A) and bluer than line (B).

\begin{figure}[t]
\begin{center}
 \includegraphics[width=3.6in]{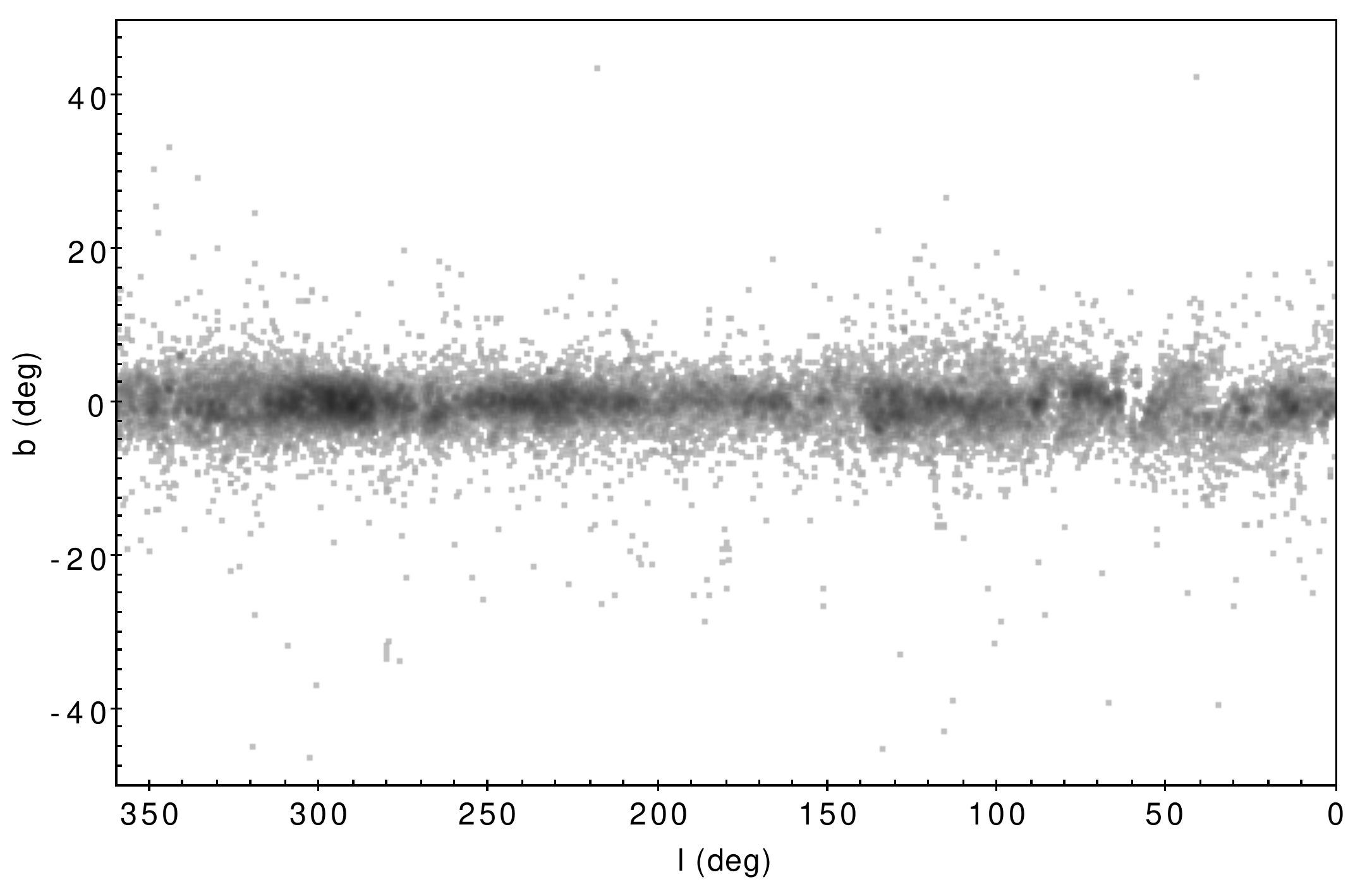}
 \caption{Location of the selected OB star candidates in the sky. The grey scale indicates stellar density.}
   \label{lb_selOB.png}
\end{center}
\end{figure}

In order to reduce the fraction of contaminants that remain in our selected region of color-color space (mostly early A
stars), we perform a second selection based on estimating the absolute magnitude of the stars using $\varpi,
\sigma_{\varpi}$ and $G$ magnitude, assuming an exponentially decreasing prior for the heliocentric distance (similar
to \cite{Bailer-Jones:2015}) and for the height from the Galactic plane. 
Taking extinction into account via the $(V - K)$ colors, we select as candidates those objects that have at least $75
\%$ probability of being brighter than B3 stars on the main sequence, resulting in $\approx$ 37000 candidate OB stars.
Figure \ref{lb_selOB.png} shows the distribution on the sky of our selected sample. Applying our selection criterium to
the Tycho-2 Spectral Type Catalogue, we estimate the amount of contamination from non-OB stars being about $40 \%$ for
stars with $\varpi < 2$ mas and $30 \%$ for $\varpi < 1$ mas. However, the presence of late OB stars (i.e. with
spectral type later than B3) is relevant ($\approx 40 \%$ of the selected stars).
The relatively high fraction of contaminants is expected to be reduced with better parallax measurements in future Gaia
releases.

{\underline{\it Future works}}. We are developing a tool aimed at determining the likelihood of a model, given an
observed sample selected as above. The objective is to perform a parameter adjustment for relevant kinematic warp
parameters, such as the warp precession or possible amplitude variations.

\bigskip
This work has made use of data from the European Space Agency (ESA)
mission {\it Gaia} (https://www.cosmos.esa.int/gaia), processed by
the {\it Gaia} Data Processing and Analysis Consortium (DPAC,
https://www.cosmos.esa.int/web/gaia/dpac/consortium). Funding for
the DPAC has been provided by national institutions, in particular the
institutions participating in the {\it Gaia} Multilateral Agreement.

\end{document}